\documentclass{cimento}

\usepackage{graphicx}  
\usepackage{amssymb} 

\title{Satellite observations of reconnection\ETC\ between emerging and pre-existing small-scale magnetic fields}
\author{S.L.~Guglielmino\from{ins:a}\ETC,
F.~Zuccarello\from{ins:a},
P.R.~Young\from{ins:b}\from{ins:c}\from{ins:d},
P.~Romano\from{ins:e}
        \atque
M.~Murabito\from{ins:f}}
\instlist{\inst{ins:a} Dipartimento di Fisica e Astronomia ``Ettore Majorana'' -- Sez.~Astrofisica, Universit\`{a} degli Studi di Catania, Via S.~Sofia 78, I-95123 Catania, Italy
	\inst{ins:b} College of Science, George Mason University, Fairfax, VA 22030, USA
	\inst{ins:c} Code 671, NASA Goddard Space Flight Center, Greenbelt, MD 20771, USA
	\inst{ins:d} Northumbria University, Newcastle upon Tyne, NE1 8ST, UK
	\inst{ins:e} INAF -- Oss.~Astrofisico di Catania, Via S.~Sofia 78, I-95123 Catania, Italy
	\inst{ins:f} INAF -- Oss.~Astronomico di Roma, via Frascati 33, I-00078 Monte Porzio Catone, Italy}

\begin{document}

\maketitle

\begin{abstract}
We report multi-wavelength ultraviolet observations taken with the \textit{IRIS} satellite, concerning the emergence phase in the upper chromosphere and transition region of an emerging flux region (EFR) embedded in the unipolar plage of active region NOAA 12529. 
The photospheric configuration of the EFR is analyzed in detail benefitting from measurements taken with the spectropolarimeter aboard the \textit{Hinode} satellite, when the EFR was fully developed. In addition, these data are complemented by full-disk, simultaneous observations of the \textit{SDO} satellite, relevant to the photosphere and the corona. In the photosphere, magnetic flux emergence signatures are recognized in the fuzzy granulation, with dark alignments between the emerging polarities, cospatial with highly inclined fields. In the upper atmospheric layers, we identify recurrent brightenings that resemble UV bursts, with counterparts in all coronal passbands. These occur at the edges of the EFR and in the region of the arch filament system (AFS) cospatial to the EFR. Jet activity is also found at chromospheric and coronal levels, near the AFS and the observed brightness enhancement sites. The analysis of the \textit{IRIS} line profiles reveals the heating of dense plasma in the low solar atmosphere and the driving of bi-directional high-velocity flows with speeds up to 100 km/s at the same locations. Furthermore, we detect a correlation between the Doppler velocity and line width of the Si IV 1394 and 1402~\AA{} line profiles in the UV burst pixels and their skewness. Comparing these findings with previous observations and numerical models, we suggest evidence of several long-lasting, small-scale magnetic reconnection episodes between the emerging bipole and the ambient field. This process leads to the cancellation of a pre-existing photospheric flux concentration of the plage with the opposite polarity flux patch of the EFR. The reconnection appears to occur higher in the atmosphere than usually observed.
 
\end{abstract}


\section{Introduction}

The advent of the \textit{IRIS} satellite \cite{DePontieu:14} provided overwhelming evidence of small-scale energy release events that occur in the solar atmosphere. In particular, \cite{Peter:14} reported on intense, transient brightenings seen in ultraviolet (UV) images. They consist of compact ($\approx 2^{\prime\prime}$) knots exhibiting a factor of about $10^{3}$ intensity increase in UV lines, with a strong line broadening indicating plasma ejections of $100 - 200 \,\mathrm{km\,s}^{-1}$. These events have been called \textit{IRIS} bombs or UV bursts \cite{Young:18}. They are usually associated with cospatial canceling opposite-polarity magnetic flux patches in the photosphere, like similar phenomena observed at optical wavelengths, e.g., Ellerman bombs. For this reason, they are likely to occur as a result of small-scale magnetic reconnection at low atmospheric heights. 

Indeed, reconnection may take place when an emerging flux region (EFR) interacts with a pre-existing flux system, such as the ambient field of an active region (AR), leading to energy release that heats locally the atmosphere and drives high-temperature plasma flows. This has been predicted in numerical simulations \cite{Yokoyama:95}, which also show that the relative orientation between the reconnecting flux systems regulates the energetics of the process \cite{Galsgaard:07}, with the external field acting as a guide for plasma ejections \cite{David:15}. These results have been observationally confirmed at granular scale as well \cite{Guglielmino:10,Ortiz:14}. 

Here, we report on recent joint observations performed by the \textit{IRIS}, \textit{Hinode} and \textit{SDO} satellites. They revealed long-lasting brigthenings, similar to UV bursts, in the upper chromosphere, transition region and corona, with simultaneous plasma ejections occurring where an EFR emerged within the plage of AR NOAA~12529 \cite{Guglielmino:18a,Guglielmino:19}.

\section{Observations and Data Analysis}

\begin{figure}[b]
	\centering
	\includegraphics[trim=60 175 185 80, clip, scale=0.3825]{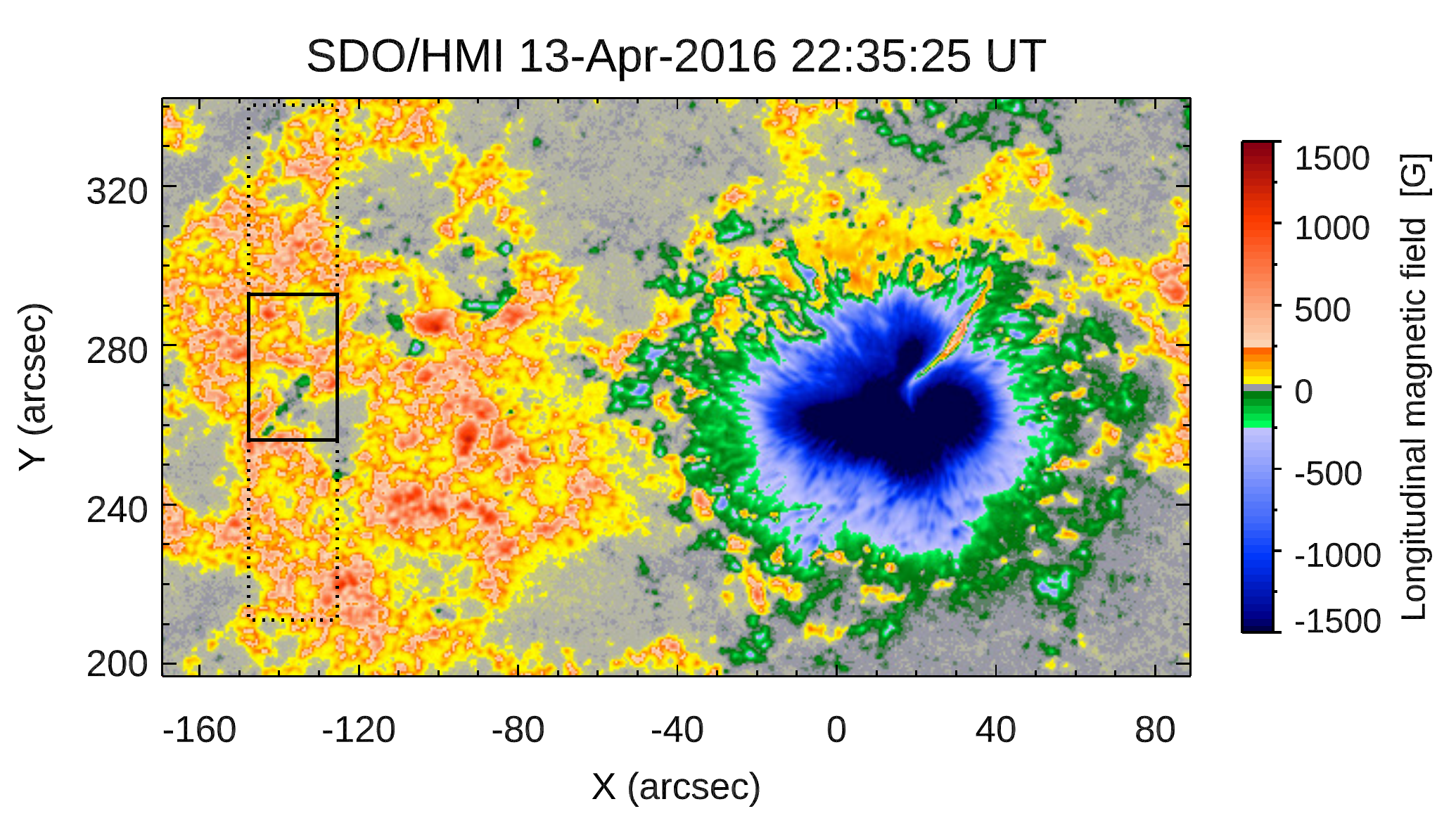}%
	\includegraphics[trim=10 10 415 265, clip, scale=0.36125]{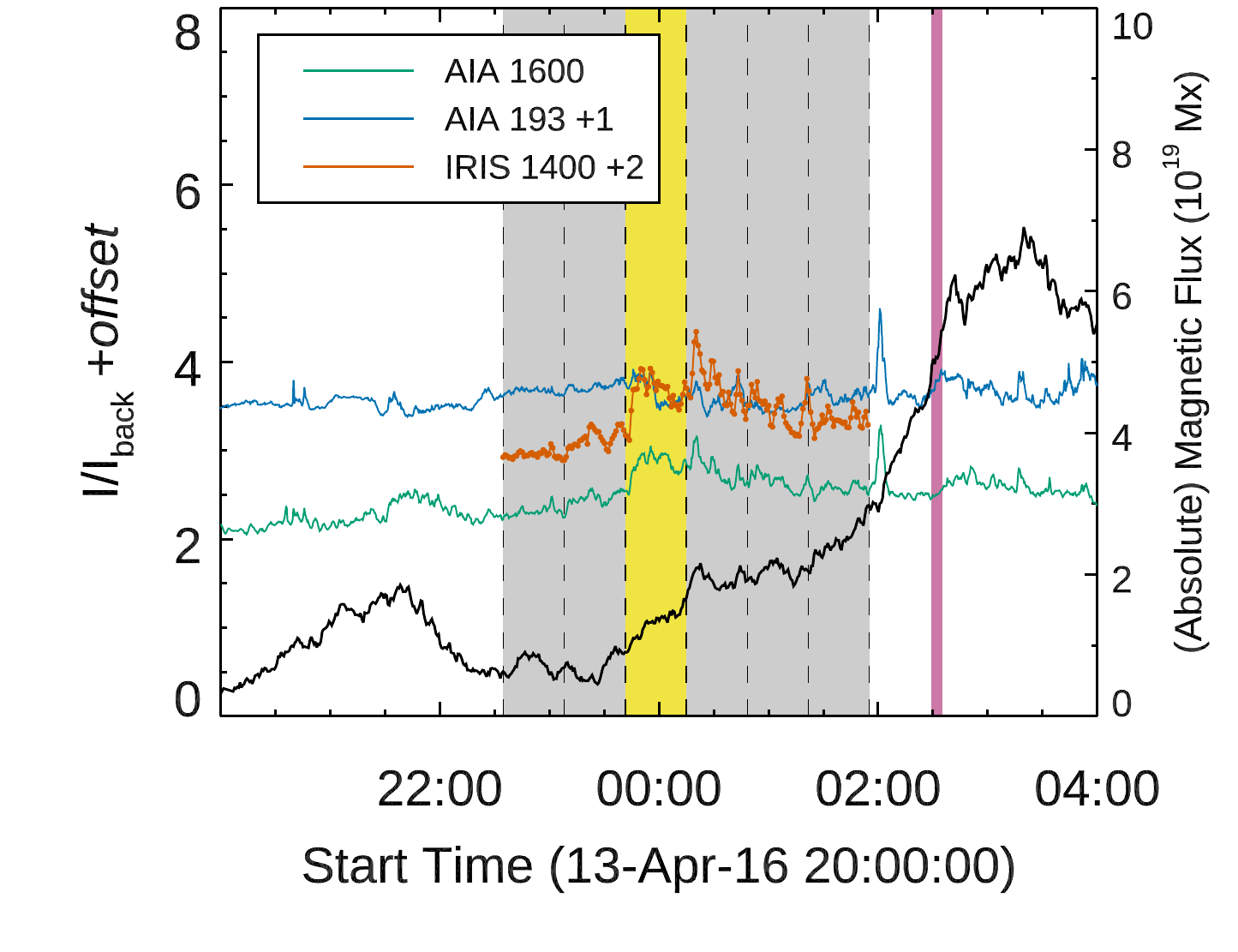}
	\caption{\textit{Left panel}: \textit{SDO}/HMI magnetogram of AR NOAA 12529. The dashed-line box represents the area scanned by the \textit{IRIS} slit. The solid-line box indicates the subFOV where the EFR appeared. \textit{Right panel}: Light curves, normalized to the background intensity of a quiet area, relative to the subFOV indicated in the magnetogram, for \textit{SDO}/AIA channels and \textit{IRIS} slit-jaw images (colors according to the legend). For comparison, the flux evolution of the EFR is also shown (black line). The grey-shaded area represents the duration of the \textit{IRIS} observing sequence. The yellow area corresponds to the \textit{IRIS} scan that is analyzed in Figure~\ref{fig:synoptic}. The magenta area indicates the time of \textit{Hinode} observations in the EFR area. \label{fig:context}}
\end{figure}

AR NOAA~12529 was observed in April 2016 \cite{Guglielmino:17}. It consisted of a large, preceding sunspot, characterized by a conspicuous umbral filament \cite{Guglielmino:17,Guglielmino:18b}, and a diffuse following polarity, with positive sign, where an EFR appeared (Figure~\ref{fig:context}, left panel). 

The flux history of the EFR (Figure~\ref{fig:context}, right panel, black line) is obtained from \textit{SDO}/HMI line-of-sight field measurements in the subFOV indicated in Figure~\ref{fig:context} (left panel). \textit{IRIS} observations caught the rising phase of the EFR. The light curve in the \textit{IRIS} 1400~\AA{} passband (Figure~\ref{fig:context}, right panel, orange line), which refers to the transition region, shows bursty brightness enhancements throughout the observing sequence. Simultaneous brightenings are observed in the \textit{SDO}/AIA 1600~\AA{} (upper photosphere) and 193~\AA{} (low corona) channels (Figure~\ref{fig:context}, right panel, green and blue lines). 

\begin{figure}[t]
	\centering
	\includegraphics[trim=10 190 205 105, clip, scale=0.55]{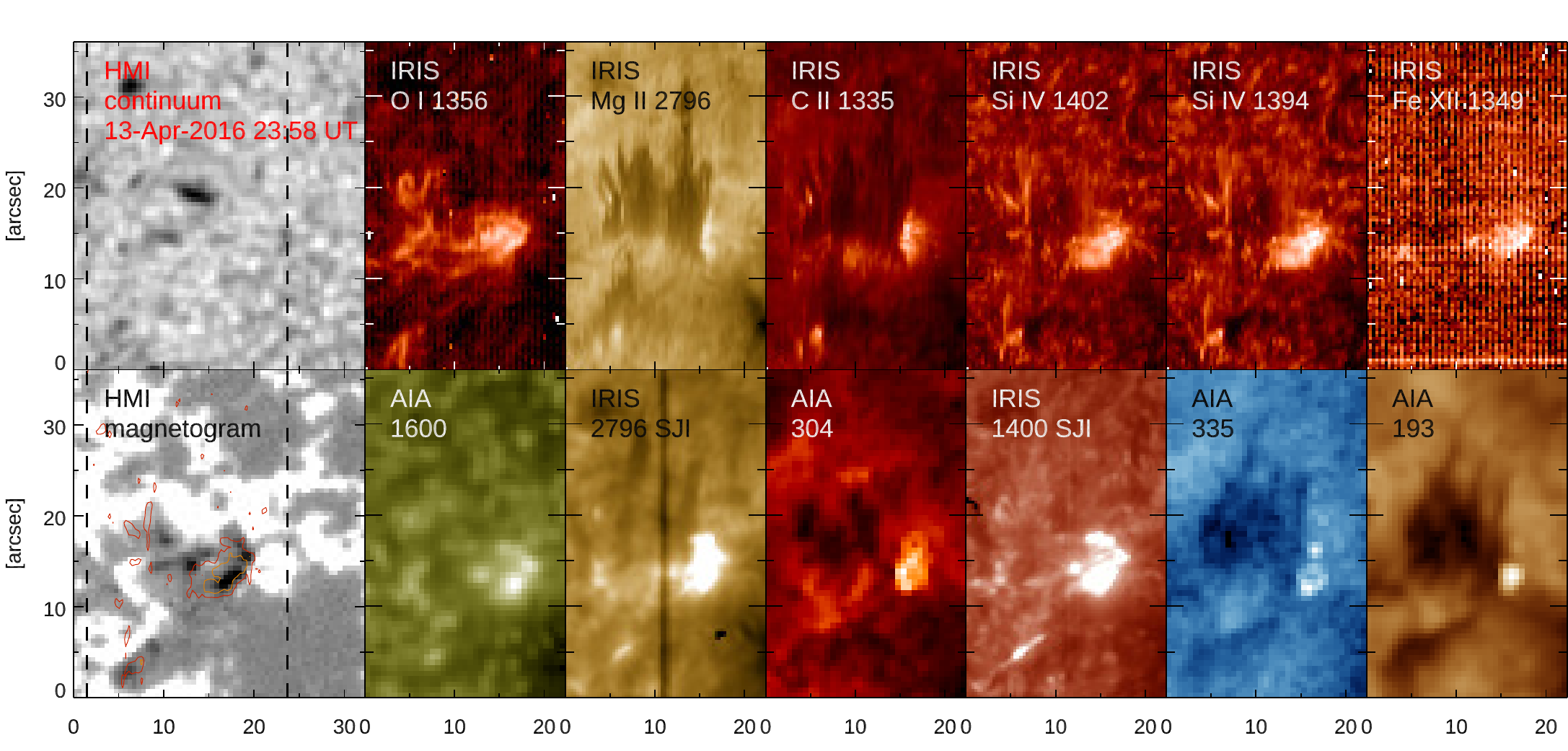}
	\caption{Synoptic view of the EFR area at different atmospheric layers, during the third \textit{IRIS} scan, from simultaneous multi-wavelength measurements of \textit{IRIS}, \textit{SDO}/HMI and \textit{SDO}/AIA. The overplotted contours on the LOS magnetogram refer to the Si IV 1402~\AA{} radiance. \label{fig:synoptic}}
\end{figure}

The sequence of \textit{SDO}/HMI observations shows that a flux cancellation event occurred in the EFR area: a pore with positive polarity, corresponding to a pre-existing magnetic flux concentration of the plage, disappeared being replaced by a new pore with negative polarity, owing to the piling up of the emerging negative field of the EFR. \textit{Hinode} spectropolarimetric measurements, performed at the EFR flux peak (Figure~\ref{fig:context}, right panel, magenta area), 
clearly confirm the emergence of new magnetic field and the formation of a new pore.
\textit{Hinode} observations also reveal the fine structure of the emerging bipole: mixed-polarity fields, with enhanced linear polarization signal, are found in the region between the opposite emerging polarities, where the granulation looks fuzzy. 

Figure~\ref{fig:synoptic} provides a synoptic view of simultaneous, multi-wavelength measurements in the EFR area. Strong, compact brightenings are seen in the \textit{IRIS} UV lines, from the chromosphere up to the coronal Fe XII 1349~\AA{} line. Cospatial brightness enhancements are observed in the \textit{IRIS} 1400 and 2796~\AA{} slit-jaw images and in all of the \textit{SDO}/AIA channels. In the middle chromosphere (O I 1356~\AA{} line), we find the development of an arch filament system (AFS) appearing in emission, which is cospatial to the serpentine field. At higher chromospheric heights (e.g., Mg II 2796~\AA{} and C II 1335~\AA{} lines), we observe that plasma ejections are adjacent to the brightenings, taking place above the AFS: those are identified as surges. Notably, the Si IV 1402~\AA{} radiance contours indicate that the areas with enhanced intensity are located near the contact region between the negative emerging polarity of the EFR and the positive, pre-existing field of the plage.

Analyzing the profiles in the brightening region, we uncovered high-velocity flows ($\approx 100 \,\mathrm{km\,s}^{-1}$) in blueshifted Mg II 2796~\AA{}, C II 1335~\AA{} and Si IV 1394/1402~\AA{} lines.

\section{Concluding remarks}

The photospheric signatures observed during the evolution of the EFR fit well with the classical scenario of flux emergence \cite{Cheung:14}. Moreover, there are strong suggestions that a flux cancellation episode occurred as a result of the interaction (reconnection) between the pre-existing field of the plage of the AR and the newly emerging EFR. This event led to energy release from the chromosphere to the corona above the EFR, as pointed out by the brightenings and plasma ejections observed in all the upper atmospheric layers.

Comparing our results with numerical MHD simulations of flux emergence \cite{Nobrega:16,Nobrega:17,Nobrega:18}, it appears that the magnetic topology overlying UV bursts determines the impact on the upper atmospheric levels. Indeed, some spectral features identified in the brightening site cospatial to the EFR indicate that the reconnection occurs higher in the atmosphere than usually observed in \textit{IRIS} bombs, thus explaining the coronal intensity enhancements. That extends the results of earlier observations of flux emergence at granular scale \cite{Ortiz:16} and needs to be further explored with future measurements at higher spatial resolution.

\acknowledgments
The authors acknowledge support by the Istituto Nazionale di Astrofisica (PRIN-INAF 2014), by the Italian MIUR-PRIN grant 2012P2HRCR on ``The active Sun and its effects on space and Earth climate'', by Space Weather Italian COmmunity (SWICO) Research Program, and by the Universit\`a degli Studi di Catania (Piano per la Ricerca Universit\`{a} di Catania 2016-2018 -- Linea di intervento~1 ``Chance''; Linea di intervento~2 ``Dotazione ordinaria''). PRY acknowledges funding from NASA grant NNX15AF48G, and he thanks ISSI Bern for supporting the International Team Meeting ``Solar UV bursts -- a new insight to magnetic reconnection''. The research leading to these results has received funding from the European Union's Horizon 2020 research and innovation programme under grant agreement no.~739500 (PRE-EST project).

\end{document}